\documentclass[pdflatex,sn-apa]{sn-jnl}% APA reference style (Behavior Research Methods)
\usepackage{graphicx}%
\usepackage{multirow}%
\usepackage{amsmath,amssymb,amsfonts}%
\usepackage{amsthm}%
\usepackage{mathrsfs}%
\usepackage[title]{appendix}%
\usepackage{xcolor}%
\usepackage{textcomp}%
\usepackage{manyfoot}%
\usepackage{booktabs}%
\usepackage{algorithm}%
\usepackage{algorithmicx}%
\usepackage{algpseudocode}%
\usepackage{listings}%
\usepackage{hyperref}

\usepackage[utf8]{inputenc}
\usepackage[T2A]{fontenc}
\theoremstyle{thmstyleone}%
\theoremstyle{thmstyletwo}%

\theoremstyle{thmstylethree}%

\begin{document}
% APA 7: three or more authors are cited as "et al." from the first citation
% (apacite implements APA 6, which spells out up to five authors the first time).
% APA 7: a space between initials (the class sets none) and no "Retrieved from" before URLs.
\renewcommand{\BPBI}{.\ }
\renewcommand{\BRetrievedFrom}{}
\shortcites{cavonius_method_1974,van_den_akker_use_1947,machikhin_attenuation_2022,woodward_supercontinuum_2009,fong_novel_2025,beliaeva_estimation_2023,machikhin_color_2023,beliaeva_2023,alfano_ultimate_2006,machikhin_acousto-optical_2018,zajtsev_colorimeter_2000,omidali_acoustooptics_2023,pang_review_2022,goutzoulis_design_1994,uchida_optical_1971,ghodrati2015suitability,robertson1978cie,gladilin2025first,shi2024multi,bongard1957substitution,macadam1942visual,witt1995cie,witt1999geometric,brown1949visual,wyszecki1971new,wyszecki1971color,alman1989performance,berns1991visual,luo1986chromaticity,kim1997influence,xu2019assessing,xu2021testing,mirjalili2019color,Huang2012,kim1997new,zhao2020color,liang2018colour,pytlarz2017close,pieri2018hitting,wright1928trichromatic,guild1931colorimetric,guild1925trichromatic,cie1987vocabulary,smith1975spectral,stockman1999cone,asano2016individual,csuti2008colour,ko2023reducing,meininger2023categorical,stiles1955interim,stiles1959npl,luo2001development,asano2020categorical,sarkar2010toward,cie2006fundamental,fechner1838scheibe,benham1894artificial,campenhausen1995hundred,spitschan2018silent,uprety2021optimizing,tominaga2021measurement,aksentsev2026homogenization,bongard1955animals,lemoan2019three,fry2019scene,uchikawa1986temporal,luo2023comprehensive,shamey2014performance,AdobeRGB1998,aksentsev2026colorimeter,metha1993calibration,brainard2002display,morovic2008color,gegenfurtner2003color,fairchild2007mean,ramanath2009minimizing,long2016observer}

\title[An acousto-optic colorimeter for color psychophysics]{An acousto-optic
colorimeter for color psychophysics}


\author*[1,2]{\fnm{Sergey} \sur{Korchagin}}\email{korchagin.sergey.97@gmail.com}

\author*[1,2]{\fnm{Georgiy} \sur{Lobarev}}\email{egor.lobarev@gmail.com}

\author[1,2]{\fnm{Pavel} \sur{Aksentsev}}

\author[1]{\fnm{Pavel} \sur{Maximov}}

\author[1]{\fnm{Maria} \sur{Gracheva}}

\author[1]{\fnm{Olga} \sur{Polschikova}}

\author[1,3,4]{\fnm{Egor} \sur{Ershov}}

\affil[1]{\orgname{Color Reproduction and Synthesis Institute},
\orgaddress{\city{Moscow}, \country{Russia}}}

\affil[2]{\orgname{Moscow Center for Advanced Studies},
\orgaddress{\city{Moscow}, \country{Russia}}}

\affil[3]{\orgname{AXXX},
\orgaddress{\city{Moscow}, \country{Russia}}}

\affil[4]{\orgname{Moscow Independent Research Institute of Artificial Intelligence},
\orgaddress{\city{Moscow}, \country{Russia}}}

\abstract{\unboldmath A complete model of human color perception requires the ability to present any color as a stimulus.
Most existing measurements were collected with physical color samples, and current laboratory generators, such as displays and multi-primary LED systems, work with a small number of fixed primaries.
Published color-difference data therefore cover about 40\% of the CIE~$xy$ chromaticity diagram, while commercial wide-gamut displays already reach two thirds of it and need difference metrics there.
We describe an acousto-optic colorimeter that lets the experimenter choose the primaries.
The device selects up to four narrow bands from a supercontinuum (white-light) laser, each with electronically set wavelength and intensity, and covers 95\% of the CIE~$xy$ chromaticity diagram.
Threshold-scale color differences, presented in alternation as in a discrimination experiment, are reproduced with a mean error of 0.08--0.12 CIEDE2000 units ($\Delta E_{00}$) at five CIE-recommended color centers.
Because one color can be built from many combinations of narrow bands, the bands are chosen for the task: the finest adjustment for color matching, the least disagreement between observers for color-difference measurement.
In simulations with 151~individual observers from a published dataset, with these bands observers see colors 0.6--1.9~$\Delta E_{00}$ from the intended one on average, against 1.2--2.5~$\Delta E_{00}$ on a calibrated wide-gamut LCD, lower at every center.
The approach applies to any narrowband source and lets experimenters use saturated colors while limiting the effect of individual differences in color vision.}




\maketitle

\section{Introduction}\label{sec:intro}

\begin{figure}[b]
\centering
\includegraphics[width=\linewidth]{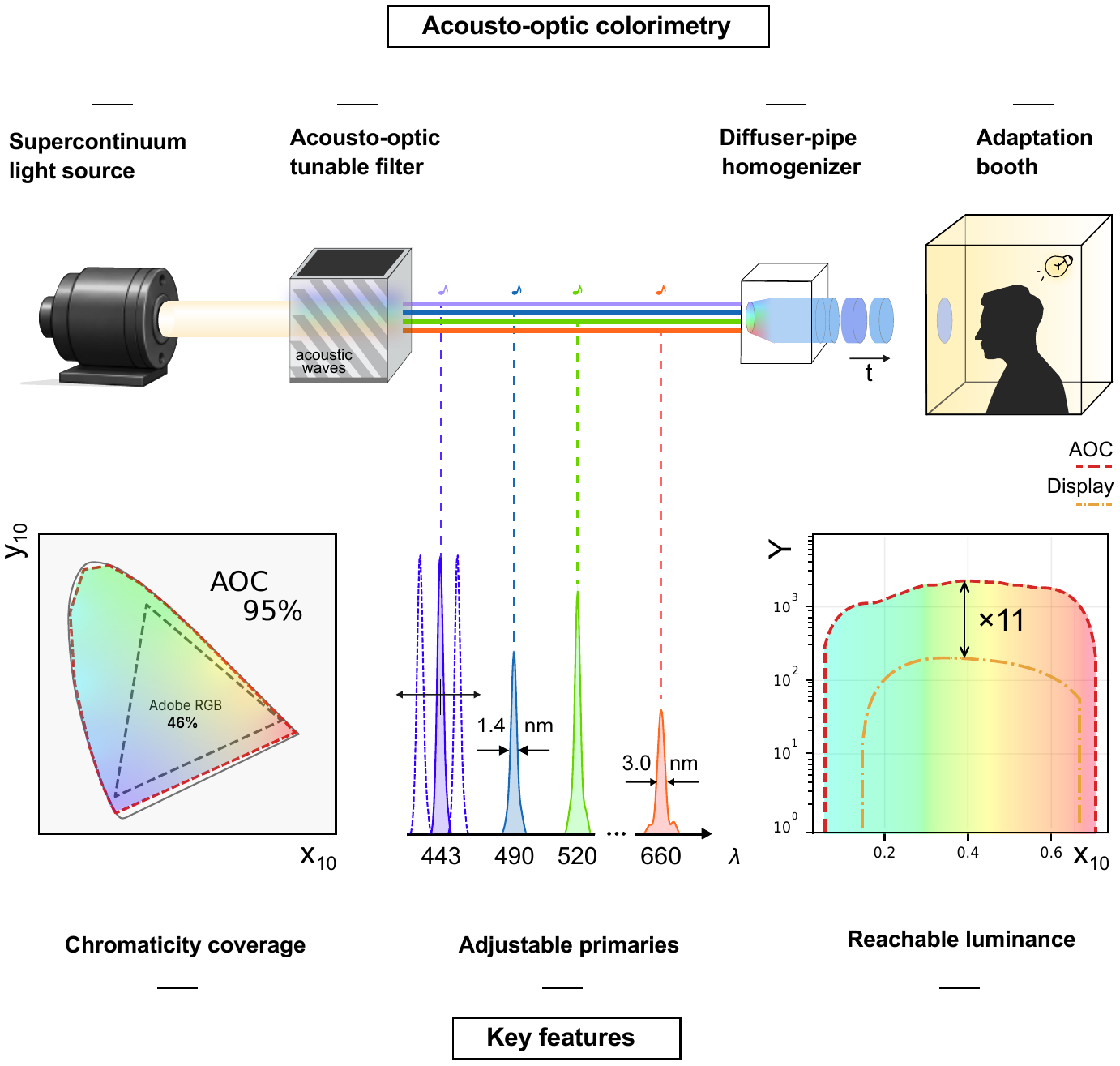}
\caption{Overview of the instrument.
\textbf{Top:} a supercontinuum light source feeds an acousto-optic tunable filter that passes up to four narrow bands, each with an electronically set wavelength and intensity (Section~\ref{sec:freedoms}). Two color stimuli are presented using the strict substitution method under fixed adaptation conditions.
\textbf{Bottom:} the stimulus is a mixture of four narrowband primaries whose peak wavelengths, not only intensities, are adjustable. The panel also compares the instrument's chromaticity coverage and reachable luminance with those of the reference display (Section~\ref{sec:instrument}).}
\label{fig:abstract}
\end{figure}

A wide range of work in vision science, engineering, and industry rests on psychophysical data about human color perception. Display calibration \citep{metha1993calibration,brainard2002display}, imaging (color-space transforms \citep{pieri2018hitting} and gamut compression \citep{morovic2008color}), color quality control \citep{alman1989performance,berns1991visual}, and visual neuroscience \citep{gegenfurtner2003color} all depend on such data. These data come from observers: the matches they set between physically different lights \citep{wright1928trichromatic,guild1931colorimetric}, the differences they can and cannot distinguish \citep{witt1995cie,luo2001development}, and the appearance they report under a given viewing condition \citep{kim1997influence}. To support reliable models and applications, psychophysical measurements must be precise and sample color space densely, across a wide gamut of chromaticities and a broad range of luminance levels.

Existing color-difference datasets provide limited coverage of highly saturated colors, particularly in regions approaching the spectral locus. The performance of color-difference formulas is therefore less well validated in these regions than in well-sampled parts of color space. Studies at wide gamut \citep{xu2021testing,zhao2020color} and high luminance \citep{xu2019assessing} have exposed disagreement between models, reinforcing the need to examine their behavior beyond conventional stimulus ranges. New psychophysical color-difference measurements near the spectral locus would provide the missing experimental evidence.

Measurement precision and coverage depend on the stimulus generator: data can be no more reliable than the accuracy, stability, and reproducibility with which stimuli are presented. Calibrated displays are stable, well characterized, and convenient, but their three broadband primaries cannot reach the most saturated colors. Generators built from a small number of narrow spectral bands are not subject to this limitation. Multi-primary LED systems \citep{shi2024multi}, laser projectors \citep{pytlarz2017close,pieri2018hitting}, and tunable-filter devices \citep{machikhin_color_2023} can reach chromaticities that no display can produce, while providing direct control over the spectrum.

The acousto-optic colorimeter (AOC) described in \citet{aksentsev2026colorimeter} is such a device (Fig.~\ref{fig:abstract}). An acousto-optic tunable filter selects up to four narrow bands from a supercontinuum laser source whose output spans the visible range in one spatially coherent beam and is smooth enough that every calibrated band is available at useful power \citep{woodward_supercontinuum_2009}. The wavelengths and intensities of the bands are set electronically. The instrument covers 95\% of the area of the spectral locus in the CIE~$xy$ diagram, against 46\% for a wide-gamut display, and reaches higher luminance (Fig.~\ref{fig:aoc_xy_coverage}). This extended gamut makes measurements near the spectral locus possible, but coverage alone does not determine whether a stimulus is suitable for a particular psychophysical task. Because the AOC uses narrowband primaries, observer metamerism must also be considered: spectra that match for the CIE standard observer may produce different colors for individual observers. The selected primary wavelengths therefore affect not only which colors the instrument can produce, but also how those colors are perceived by the observers being studied.

This paper takes the AOC from instrument characterization to use in color psychophysics and addresses two related questions: can it reproduce threshold-scale color differences, and how should its primaries be selected for different psychophysical tasks?

The earlier characterization \citep{aksentsev2026colorimeter} showed that the instrument produces a single specified color accurately and stably: with feedback stabilization enabled, it reproduces the five CIE color centers to within 0.34 to 1.26~$\Delta E_{00}$ over ten-minute intervals, in a field whose mean spatial non-uniformity is 0.91~$\Delta E_{00}$. A color-difference experiment, however, depends on the difference between two stimuli presented in alternation, and that difference must be reproduced accurately at the scale being judged, including values near discrimination thresholds. Section~\ref{sec:approbation} measures this directly: set color differences of 0.4 to 3.7~$\Delta E_{00}$ are realized with a mean absolute error of 0.08 to 0.12~$\Delta E_{00}$ under continuous alternation over a 54-hour experiment, and the instrument is compared with a calibrated display using the same protocol. This validation of reproduced color differences is the first contribution of the paper.

The second contribution is a task-oriented method for selecting the AOC's primary wavelengths. The same color can be produced with many different sets of primaries; within the device model, every such set matches the target exactly for the CIE standard observer, so colorimetric accuracy alone cannot distinguish among them. For individual observers, however, spectra that match for the standard observer may appear different. This observer metamerism is especially relevant to narrowband generators: a narrow primary samples an observer's color-matching functions at essentially one wavelength, whereas a broad primary averages over a band. The degree of departure from the specified color therefore depends on the selected primaries. Display studies have examined primary selection with observer metamerism in mind \citep{fairchild2007mean,ramanath2009minimizing,long2016observer}, but a display's primaries are fixed once it is built. The AOC allows its primary wavelengths to be selected electronically, for each target and according to the task. For measuring individual color-matching functions by the method of adjustment, the criterion is the finest color resolution, or the smallest color step the instrument can produce, because this step limits how finely an observer can set a match. For color-difference measurements, the criterion is the least observer metamerism. These criteria select different primaries and lead to substantially different predictions for individual observers. Evaluated for the 151 color-normal observers of an individual-observer model \citep{asano2016individual}, primaries selected for the finest color resolution place observers 1.9 to 4.9~$\Delta E_{00}$ from the specified color on average, more than a calibrated display (1.2 to 2.5~$\Delta E_{00}$) at every color center. On the same instrument, primaries selected for the least observer metamerism place them 0.6 to 1.9~$\Delta E_{00}$ away, closer than the display at every center. Together, these results establish the central principle of this work: a tunable narrowband colorimeter should be configured for the psychophysical task, not merely for colorimetric accuracy.

Further setups, including flicker-induced colors, silent substitution, observer classification, and objective colorimetry on animals, are outlined more briefly as proposals. None of these selections costs light, hardware, or reproduction accuracy, since each is a computation over calibration data already in hand. The argument is not specific to acousto-optics and applies to any generator built from a small number of narrow primaries.

Section~\ref{sec:related} reviews prior work on stimulus generation for color-difference research, on the limits of the standard observer, and on models of individual observers. Section~\ref{sec:instrument} summarizes the AOC as far as this paper uses it, and Section~\ref{sec:freedoms} describes the degrees of freedom of its configuration. Section~\ref{sec:approbation} summarizes the basic characterization reported in \citet{aksentsev2026colorimeter} and reports the validation of reproduced color differences. Section~\ref{sec:primary_selection} selects the configuration for each task and compares the resulting stimuli with those of a reference display. Section~\ref{sec:discussion} discusses the implications for users of comparable instruments and the limitations of the analysis, and Section~\ref{sec:conclusion} concludes.
\section{Related work}\label{sec:related}

Three lines of prior work converge in this paper.
The first concerns the generation of stimuli for color-difference experiments. 
The second addresses observer metamerism: two stimuli that match for the CIE standard observer may not match for an individual observer, particularly when narrowband spectra are involved. 
The third concerns the models of individual observers, which turn this known effect into a quantity that can be computed for any spectrum before an experiment.
The acousto-optic colorimeter belongs to the first strand, is exposed to the second, and, because its primaries are selected electronically, can use the third to choose its stimuli.

The experimental basis of color-difference metrics consists primarily of datasets obtained using physical color samples.
The measurements of \citet{macadam1942visual} demonstrated the perceptual nonuniformity of the color spaces available at that time \citep{cie1987vocabulary}, and later studies confirmed it \citep{brown1949visual,wyszecki1971new,wyszecki1971color}.
The datasets compiled in COMBVD, including RIT--DuPont \citep{alman1989performance,berns1991visual}, BFD-P \citep{luo1986chromaticity}, Witt \citep{witt1995cie,witt1999geometric} and Leeds \citep{kim1997influence,kim1997new}, remain the standard for evaluating color-difference formulas.
They and several more recent collections \citep{xu2019assessing,mirjalili2019color,Huang2012} were built from colored patches, whose chromaticities are limited by the achievable range of pigments.
Consequently, the available experimental data provide dense coverage of the central region of the chromaticity diagram but remain sparse near the spectral locus (Fig.~\ref{fig:aoc_xy_coverage}a).
Among the available datasets, the measurements of \citet{macadam1942visual} and the display-based dataset of \citet{zhao2020color} extend furthest toward the spectral locus, although the latter remains constrained by the DCI-P3 display gamut.
The available datasets therefore provide only limited validation of color-difference formulas in highly saturated regions. 
A comprehensive evaluation against existing datasets showed that their performance varies across different regions of color space \citep{luo2023comprehensive}.
This dependence becomes particularly important at high chroma: measurements at a saturated blue color center revealed substantial differences in both the accuracy and relative ranking of the formulas \citep{shamey2014performance}.
These results indicate that color-difference formulas require dedicated experimental validation in regions approaching the spectral locus, which remain sparsely sampled by existing datasets.
The gap matters in practice, because wide-gamut and high-dynamic-range displays now reproduce colors outside the well-sampled region of color space \citep{pytlarz2017close,pieri2018hitting}: together the COMBVD sets span 39\% of the area of the spectral locus and MacAdam's centers 48\%, whereas a DCI-P3 gamut spans 48\% and a Rec.~2020 gamut 67\% (Fig.~\ref{fig:aoc_xy_coverage}a).

Closing this gap requires both a suitable instrument and a protocol for configuring and using it in color-difference experiments. This work validates the AOC for threshold-scale color-difference measurements and develops a protocol for selecting its primaries according to the psychophysical task.
To measure color differences at a saturated chromaticity, down to discrimination thresholds, a generator has to reach that chromaticity, resolve steps well below the smallest difference measured, and alternate the two stimuli of a pair on the same retinal area \citep{gladilin2025first}, all stably over a session.
Calibrated displays are stable and flexible \citep{xu2021testing,zhao2020color,liang2018colour,xu2019assessing,ghodrati2015suitability}, but their three broadband primaries cannot reach most saturated colors.
Generators built from narrow bands can.
They range from the monochromators of the classical matching experiments \citep{guild1925trichromatic,guild1931colorimetric,wright1928trichromatic,bongard1957substitution} to laser projectors \citep{pytlarz2017close,pieri2018hitting} and multi-primary LED systems \citep{shi2024multi}, each trading spectral precision, speed and intensity differently.
An acousto-optic tunable filter combined with a supercontinuum source provides electronically selectable narrow bands at useful luminance, with no moving parts \citep{machikhin_acousto-optical_2018,machikhin_color_2023}.
The instrument used in this paper was described and characterized by \citet{aksentsev2026colorimeter}.

Narrowband generators share a problem that concerns observers rather than photometry.
Colorimetry treats two stimuli with equal tristimulus values for the standard observer as interchangeable.
That observer, however, is an average of a few dozen people \citep{wright1928trichromatic,guild1931colorimetric,stiles1955interim,stiles1959npl}, and individuals differ from it in lens and macular pigment density, photopigment density and the peak sensitivities of their cones, which the cone fundamentals make explicit \citep{smith1975spectral,stockman1999cone,cie2006fundamental}.
A stimulus that matches the specified color for the standard observer can therefore depart from it for a given person.
% \todo{Можем ли писать <<We call>>? Это вроде общепринято?}
%https://en.wikipedia.org/wiki/Metamerism_(color)
This effect is known as \emph{observer metamerism} \citep{fairchild2007mean}, and it is the quantity this paper minimizes.
Broad spectra average individual differences over many wavelengths, but a narrow line samples an observer's color-matching functions at essentially one wavelength, so narrowband stimuli are affected most.
Matches made with narrowband LED primaries depart measurably from the standard observer's prediction \citep{csuti2008colour}, and visually matched whites on an LCD and a laser projector differed by up to 18.5~$\Delta E_{00}$ \citep{ko2023reducing}.

Measuring color-matching functions individually is laborious, and only a few studies have done so.
The individual colorimetric observer model \citep{asano2016individual} fits physiological parameters to each observer's matching data and reconstructs that observer's color-matching functions, which provides a population of 151 color-normal observers against which any spectrum can be evaluated.
The population is conservative: it contains only the variation the model parameters can represent, and parameter bounds compress its tails.
The departures reported in this paper are therefore lower estimates (Section~\ref{subsec:lim_observers}).

These lines of research have been connected for displays, but less so for stimulus generators.
Narrowband generators are qualified by accuracy, stability and uniformity \citep{shi2024multi,aksentsev2026colorimeter}, criteria defined entirely for the standard observer.
Where observer metamerism has been addressed, most remedies change the observer description rather than the stimulus: cone fundamentals in place of the standard observer \citep{csuti2008colour}, fundamentals refitted to observers' matches \citep{ko2023reducing}, and categorical observers \citep{sarkar2010toward,asano2020categorical,meininger2023categorical}.
On the stimulus side, display studies have shown that observer metamerism grows as primaries narrow and have examined the choice of display primaries with it in mind \citep{fairchild2007mean,ramanath2009minimizing}, and multiprimary displays have been proposed to reduce it \citep{long2016observer}.
These analyses inform the design of a device whose primaries are then fixed in manufacture, so once a display is built, the observer description is the only term left free.
A narrowband generator with electronically selectable primaries can make the choice of primaries at run time instead, separately for each target and each experiment, from calibration data already in hand.
To our knowledge, this has not been done for a stimulus generator used in color psychophysics.
Section~\ref{sec:primary_selection} shows that on such a generator the feasible primary sets differ in observers' mean departure from the specified color by a factor of three to eight, at no cost in reproduction accuracy.
\section{Acousto-optic colorimeter}\label{sec:instrument}

\begin{figure}[tp]
    \centering
    \includegraphics[width=1.0\linewidth]{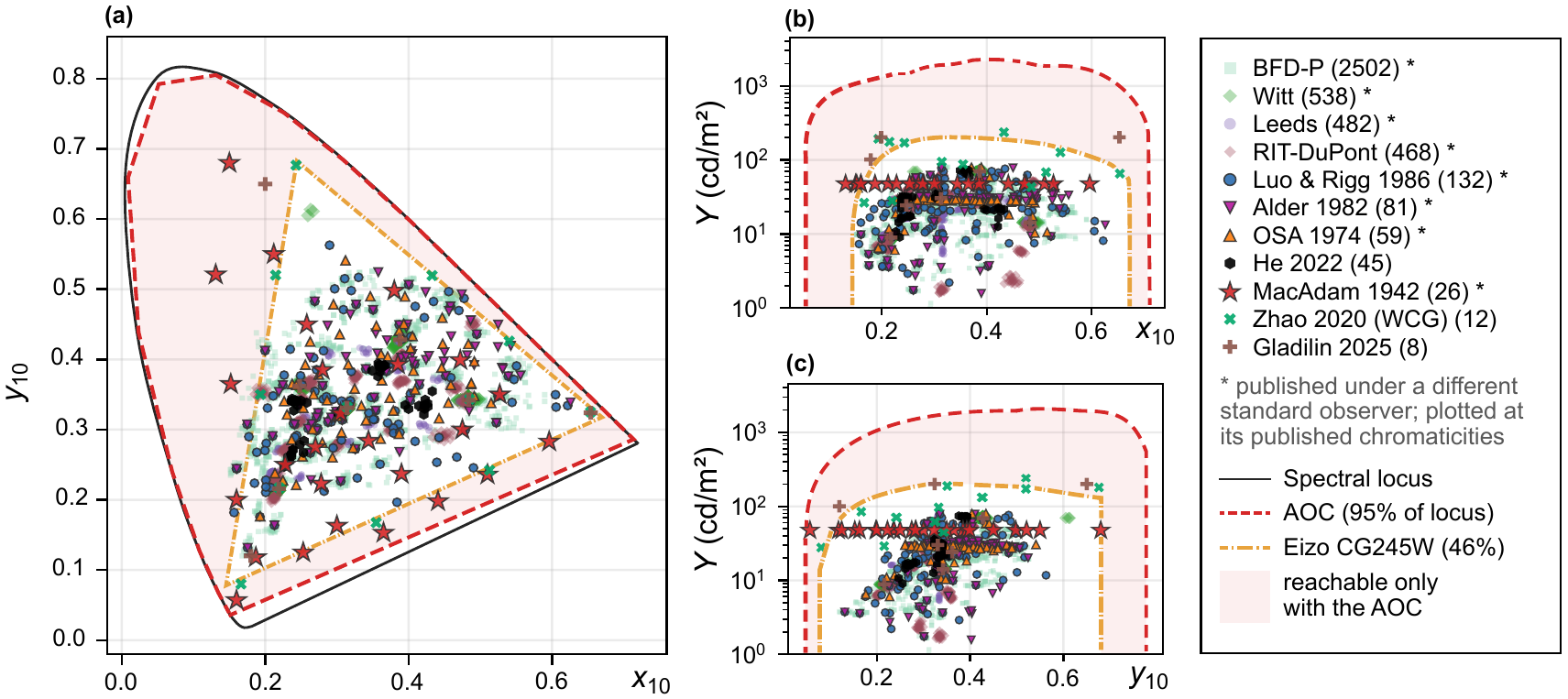}
    \caption{Coverage of the acousto-optic colorimeter (AOC) and of the reference display, with the color centers of published color-difference datasets overlaid.
(a) Chromaticity on the CIE 1964 $10^\circ$ $xy$ diagram. Sets marked $*$ are published under a different standard observer, several of them under illuminant~C, and are plotted at their published coordinates.
The AOC boundary is computed from the spectra of the feedback spectroradiometer, which resolve the bands more finely than the X-Rite calibration.
(b,~c) The reachable color solid projected on $x$ and on $y$, on a logarithmic luminance axis, for every four-primary subset of the nineteen calibrated wavelengths at its per-channel ceilings. The display's solid is the additive mixture of its three measured primaries.
Surface datasets are plotted at their published relative $Y$ against the 100~cd/m$^2$ adaptation level, and emissive sets at the luminances measured by their authors.}
    \label{fig:aoc_xy_coverage}
\end{figure}

The instrument under study produces a color by adding together up to four narrow spectral bands.
A supercontinuum source whose filtered output covers 380--750~nm illuminates a four-channel acousto-optic tunable filter.
Each channel is driven by a single acoustic frequency--amplitude pair $(f,u)$: the frequency selects the wavelength of the band that the channel diffracts, the amplitude selects its intensity \citep{machikhin_color_2023}.

The bands leave the filter at different angles, separated over an output aperture of about 8~mm, so the stimulus has to be formed rather than simply emitted.
A homogenizer of two diffuser-and-light-pipe stages in series mixes the bands, expands the beam and projects it onto a continuously rotating diffusive screen \citep{aksentsev2026homogenization}.
The source is coherent, and the rotation replaces the static speckle that a stationary diffuser would lay over the stimulus with a succession of decorrelated realizations, which the visual system averages over its own integration time.
The observer views a circular field 18~mm in diameter, which subtends $2^\circ$ at a viewing distance of 515~mm.

The total power of the supercontinuum varies by about 10\% around its mean on a timescale of half an hour, so the output is stabilized in a closed loop.
A feedback spectroradiometer (YIXIST YOA-8404-05) reads a fraction of the beam picked off the uncoated first surface of the homogenizer, and a proportional controller holds each channel's integrated line power at the value the target requires, updated at 50~Hz \citep{aksentsev2026colorimeter}.
The loop was enabled continuously throughout every measurement reported here, and Section~\ref{sec:approbation} measures what it achieves.
The calibration used throughout was measured after two hours of warm-up of the supercontinuum source, by stepping the drive of each band and channel through its range in 1\% increments and recording the spectrum at every level \citep{aksentsev2026colorimeter}.

The optical design, calibration and basic performance of the colorimeter are reported by \citet{aksentsev2026colorimeter}.
This section summarizes the properties on which the present paper relies.

The arrangement gives the instrument the following properties:

\begin{enumerate}
\item The bands are 1.4 to 3.2~nm wide at half maximum across the calibrated range \citep{aksentsev2026colorimeter}, which gives a gamut far wider than a display's: 95\% of the area of the spectral locus in chromaticity, against 46\% for a hardware-calibrated EIZO CG245W LCD monitor with Adobe RGB color space \citep{AdobeRGB1998}, and a higher attainable luminance (Fig.~\ref{fig:aoc_xy_coverage}).

\item The wavelengths are not fixed.
The drive frequency is continuous over the 66--140~MHz operating range \citep{aksentsev2026colorimeter}, so the selectable bands form a continuum, of which the calibration used in this work samples nineteen wavelengths, from 443 to 660~nm, over the 76--130~MHz of drive that this span occupies.
Each of them can be set with a precision of 0.01~MHz.
Any combination of them may serve as primaries, so the same color can be produced in many spectrally different ways -- a freedom Section~\ref{sec:freedoms} sets out and Section~\ref{sec:primary_selection} exploits.
The grid belongs to the calibration and not to the hardware: a denser one could only enlarge the choice available.

\item Each amplitude is set in steps of 0.1\% of full scale, the least significant bit of the 10-bit command, which gives about 1000 distinct levels per primary. This step is the drive quantum used in Section~\ref{sec:primary_selection}.

\item The four channels are not interchangeable.
The output of the RF driver falls by about an order of magnitude from the low to the high end of the tuning range, and the diffraction efficiency varies independently of it, with a pronounced minimum at short wavelengths \citep{aksentsev2026colorimeter}, so the intensity available depends strongly on the wavelength.
It also depends on the channel: in the calibration used in Section~\ref{sec:primary_selection}, channels 1 and 2 deliver 37 to 100\,\% of the intensity of channels 3 and 4 at the same wavelength, typically two thirds.
Each channel therefore has its own ceiling on the intensity it can deliver at a given wavelength, and each is calibrated separately.

\item The whole composition is switched by the acoustic drive alone, so stimuli can be alternated -- and the primaries themselves changed between presentations -- at rates up to 1~kHz.
No two active channels may be given the same frequency.
\end{enumerate}

The calibration that maps each frequency--amplitude pair to a radiometrically correct spectrum was made with an X-Rite i1~Pro~2, used in emission mode as a spectroradiometer, which reads a single spot at the center of the field \citep{aksentsev2026colorimeter}.
It is the radiometric reference of the instrument, and its spectra are used throughout this paper, including the observer computations of Section~\ref{sec:primary_selection}.
Its spectral resolution, about 10~nm, is coarser than the bands, so it records each band broadened, with its integrated power preserved.
Because the same spectroradiometer calibrates the colorimeter and verifies it in Section~\ref{sec:approbation}, those results are self-consistent and do not depend on the broadening.
In Section~\ref{sec:primary_selection}, a broadened band averages each observer's color-matching functions over a wider interval than the eye actually receives, which tends to understate observer metamerism rather than overstate it.
The feedback spectroradiometer resolves the bands more finely, but its spectra need additional processing to be placed on a radiometric scale.
They are used only for the AOC boundary in Fig.~\ref{fig:aoc_xy_coverage}a and the coverage derived from it.
Chromaticity does not depend on the radiometric scale, and the finer spectra place each band where it lies on the diagram, whereas the broadened X-Rite spectra would pull the boundary away from the spectral locus.
\section{Degrees of freedom of the configuration}\label{sec:freedoms}

\begin{figure}[tp]
\centering
\includegraphics[width=\linewidth]{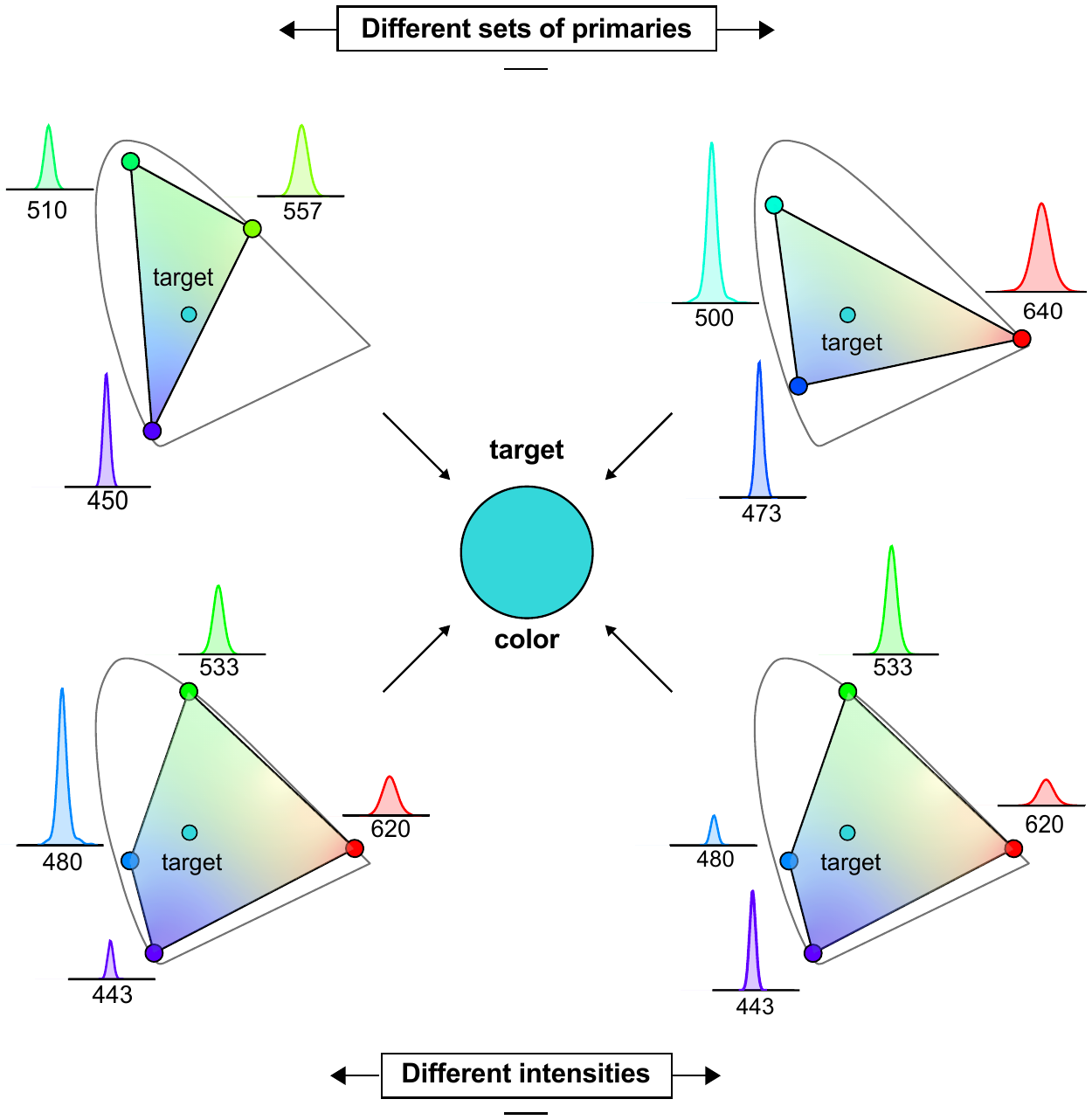}
\caption{The two free choices of the configuration, on the CIE 1964 $10^\circ$ chromaticity diagram.
\textbf{Top}: one target color reproduced by different sets of 3 primaries, each shown with its band spectra and the gamut triangle it spans.
\textbf{Bottom}: one 4 primary set reproducing the same target with different divisions of intensity among its bands, two members of the family of Eq.~\eqref{eq:ps_coef4}.}\label{fig:freedom_of_configuration}
\end{figure}

A target color can be reproduced by the AOC in many different ways, all of which are identical for the CIE standard observer.
Two choices are left free: which calibrated wavelengths serve as primaries, and, when four primaries are used, how the intensity is divided among them.
Figure~\ref{fig:freedom_of_configuration} illustrates both, and Section~\ref{sec:primary_selection} fixes them for each task.

The first choice is the set of primaries.
Any three or four of the calibrated wavelengths can be combined, as long as the target lies within the gamut they span.
A target in the middle of the chromaticity diagram can therefore be produced by hundreds of different sets, with very different spectra.

The second choice is the division of intensity among the primaries.
A mixture of the calibrated primary spectra $\Vec{S}_{i}$ of a set $\Lambda$, with intensities $c_{i}$, reproduces a target with tristimulus values $\Vec{t}$ for the CIE 1964 $10^\circ$ standard observer $\bar{T}$ when $A_{\Lambda}\Vec{c}=\Vec{t}$.
Here $\Vec{c}=(c_{i})_{i\in\Lambda}$, and the columns of $A_{\Lambda}=\bigl[\bar{T}\Vec{S}_{i}\bigr]_{i\in\Lambda}$ are the tristimulus values of the primaries.
With three primaries this system has exactly one solution,
\begin{equation}\label{eq:ps_coef3}
\Vec{c}\;=\;A_{\Lambda}^{-1}\,\Vec{t},
\end{equation}
and the set can be used only if all three intensities are non-negative.
With four primaries there are three equations for four unknowns, so the solutions form a one-parameter family,
\begin{equation}\label{eq:ps_coef4}
\Vec{c}(\tau)\;=\;\Vec{c}_{0}+\tau\,\Vec{n},\qquad A_{\Lambda}\,\Vec{n}=\Vec{0},\qquad \tau\in[\tau^{-},\tau^{+}],
\end{equation}
where $\Vec{c}_{0}$ is one solution, $\Vec{n}$ is a direction along which intensity can be moved between the primaries without changing the color for the standard observer, and $[\tau^{-},\tau^{+}]$ is the range over which all intensities stay non-negative.
Every member of the family has the same color for the standard observer, but a different spectrum, and therefore possibly a different color for an individual observer.

In practice the intensities are computed from the calibration of the instrument, following \citet{aksentsev2026colorimeter}.
Because the shape of each band changes slightly with drive level, the computation reads each primary's measured spectrum at its current intensity and iterates until the result settles.
With four primaries it also sweeps the family of Eq.~\eqref{eq:ps_coef4} and picks one member by the rule of the task (Section~\ref{sec:primary_selection}).
A set is rejected if any primary would need more intensity than its channel can deliver, and the final drive is rounded to the drive quantum of 0.1\%.
Every configuration in this paper is computed this way, so each one can actually be produced by the instrument.

Finally, the drive can switch the whole spectral composition electronically, with a period as short as 1~ms.
Unlike the two choices above, this does not leave the stimulus undetermined but changes it, and its setting follows from the experimental design.
Because the visual system integrates chromatic changes over roughly 100 to 200~ms \citep{uchikawa1986temporal}, fast switching between bands is seen as their average, which in principle allows broader spectra to be produced as well (Section~\ref{subsec:task_fechner}).
\section{Validation of the acousto-optic colorimeter}\label{sec:approbation}

The basic performance of the instrument was characterized by \citet{aksentsev2026colorimeter}.
Twelve hours after calibration, with feedback stabilization enabled, it reproduced the five CIE color centers used in this paper with mean errors of 0.34 to 1.26~$\Delta E_{00}$ over ten-minute intervals.
Over the 18~mm field the mean spatial color difference was 0.91~$\Delta E_{00}$, dominated by a gradient in luminance rather than in spectral shape \citep{aksentsev2026colorimeter,aksentsev2026homogenization}.
A color-difference experiment needs more than this.
It presents two stimuli in alternation and depends on the difference between them, so this difference must be reproduced accurately at the scale being judged.
This section reports two further results: how accurately set color differences are reproduced, and how the instrument compares with a calibrated display measured in the same way.

\begin{table}[h]
\caption{Reproduction of the color centers and of the set color differences by the colorimeter and, for comparison, by the reference display measured with the same protocol.
The error at the center is the $\Delta E_{00}$ between the measured and target colors.
The pair-separation error is the absolute deviation of the measured from the set color difference.
Values are mean $\pm$ SD over all readings of each center (the colorimeter over its fourteen blocks, the display over its two cycles), and the better value of each pair of columns is shown in bold.
The colorimeter primaries are the three-primary sets selected by the resolution criterion of Section~\ref{sec:primary_selection}, as used in operation.
All values in $\Delta E_{00}$.}
\label{tab:accuracy}%
\begin{tabular*}{\textwidth}{@{\extracolsep\fill}lccccc}
\toprule%
 & Colorimeter primaries & \multicolumn{2}{c}{Error at the center} & \multicolumn{2}{c}{Pair-separation error} \\
\cmidrule(lr){3-4}\cmidrule(lr){5-6}
Color center & (nm) & colorimeter & display & colorimeter & display \\
\midrule
Gray   & 457, 480, 573 & $\mathbf{0.61 \pm 0.26}$ & $0.73 \pm 0.04$ & $0.11 \pm 0.09$ & $\mathbf{0.07 \pm 0.05}$ \\
Yellow & 457, 510, 573 & $\mathbf{0.59 \pm 0.25}$ & $0.74 \pm 0.05$ & $\mathbf{0.09 \pm 0.07}$ & $\mathbf{0.09 \pm 0.07}$ \\
Green  & 457, 490, 603 & $0.88 \pm 0.26$ & $\mathbf{0.81 \pm 0.05}$ & $0.12 \pm 0.11$ & $\mathbf{0.05 \pm 0.05}$ \\
Blue   & 450, 473, 573 & $0.41 \pm 0.15$ & $\mathbf{0.18 \pm 0.03}$ & $0.09 \pm 0.06$ & $\mathbf{0.08 \pm 0.06}$ \\
Red    & 450, 500, 603 & $\mathbf{0.48 \pm 0.22}$ & $0.69 \pm 0.04$ & $\mathbf{0.08 \pm 0.07}$ & $0.09 \pm 0.07$ \\
\midrule
Mean   &               & \textbf{0.59} & 0.63 & 0.10 & \textbf{0.08} \\
\botrule
\end{tabular*}
\end{table}

\label{subsec:accuracy}%
An X-Rite i1~Pro~2 spectroradiometer was placed at the observer's position and aimed at the center of the field.
The stimuli were measured at the five color centers recommended by the CIE for color-difference research \citep{robertson1978cie}, which are used throughout this paper.
% \todo[inline]{"The instrument was measured" - maybe not an instrument after all?}
In CIE 1964 $10^\circ$ color coordinates they are gray $(0.314, 0.331)$ at 30~cd/m$^2$, yellow $(0.388, 0.428)$ at 69.3~cd/m$^2$, green $(0.248, 0.362)$ at 24.0~cd/m$^2$, blue $(0.219, 0.216)$ at 8.8~cd/m$^2$ and red $(0.484, 0.342)$ at 14.1~cd/m$^2$.
At each center, the instrument alternated between two stimuli: the center itself and a test color displaced from it in one of eight directions in color space, by one of five set steps of 0.5 to 2.5~$\Delta E_{00}$.
Five separations in eight directions give forty pairs per center.
The forty pairs were measured in turn, each pair alternated at 5~Hz before its reading; one pass through the forty, called a \emph{sweep}, takes two to three and a half minutes.
Each center was measured in fourteen \emph{blocks} of 24~min, the five centers alternating over a 54-hour experiment.
All blocks were measured with feedback stabilization enabled throughout.
All color differences in this section are computed as $\Delta E_{00}$ \citep{luo2001development} for the CIE 1964 $10^\circ$ standard observer with a D50 white point at 100~cd/m$^2$.

\begin{figure}[h]
\centering
\includegraphics[width=\linewidth]{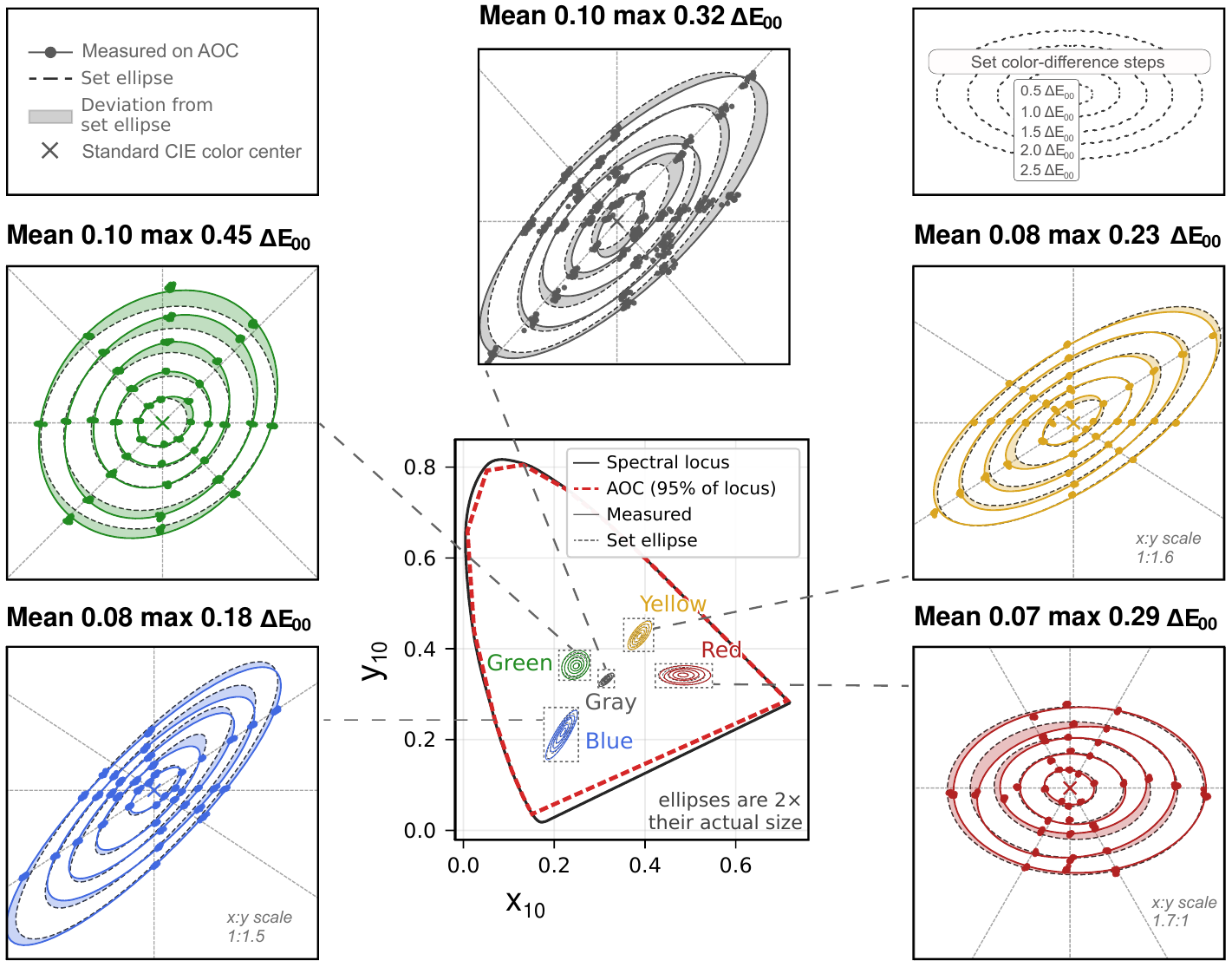}
\caption{Reproduction of the set stimulus pairs at the five color centers, from the 54-hour experiment of Table~\ref{tab:accuracy}.
\textbf{Center}: the colorimeter's coverage on the CIE 1964 $10^\circ$ diagram, each center's pair structure drawn in place at twice actual size.
\textbf{Panels}: one color center each, scaled to the window its five set steps occupy.
Markers show the measured test stimulus in the eight protocol directions, one per measurement block.
The header above each panel gives the deviation of the measured from the set color difference, averaged over the readings of each pair.
The solid contour is the ellipse fitted to the markers, the dashed contour is the ellipse through the set test colors.
}\label{fig:fidelity}
\end{figure}

The same protocol was applied to a reference display, which serves throughout this paper as the baseline for the colorimeter.
The display is an EIZO CG245W hardware-calibrated LCD monitor (1920~$\times$~1200 pixels, 10 bits per channel raised to an effective 12 bits by ordered spatial dithering, maximum luminance 180~cd/m$^2$, gamut covering Adobe~RGB, Fig.~\ref{fig:aoc_xy_coverage}), a model family used as a stimulus generator in color-difference and other psychophysical studies \citep{liang2018colour,lemoan2019three,fry2019scene}, whose calibration and stability are well studied \citep{ghodrati2015suitability}.
After warm-up it was calibrated in hardware to a gamma of 1.0, its primaries and its black level (0.39~cd/m$^2$) were measured with the X-Rite i1~Pro~2 spectroradiometer, and all stimuli were converted to RGB through the resulting display model, linear in the primaries with the black level added.
Both devices were set to reproduce the same nominal targets.
The display was measured at each center in two cycles of about an hour, which gives 36 readings of every pair.

Table~\ref{tab:accuracy} summarizes both results.
At the color centers, the colorimeter holds its output within 0.41 to 0.88~$\Delta E_{00}$ of the target over the whole experiment, and the display measured the same way within 0.18 to 0.81~$\Delta E_{00}$: the colorimeter is closer at gray, yellow and red, the display at green and blue, and the two are level on average (0.59 against 0.63).
The set color differences are reproduced with a mean absolute deviation of 0.08 to 0.12~$\Delta E_{00}$, depending on the center, with a maximum of 1.40 for a single reading.
This is comparable to the display's 0.05 to 0.09~$\Delta E_{00}$, slightly higher on average (0.10 against 0.08): the display is closer at gray, green and blue, the two are level at yellow, and the colorimeter is closer at red.
The deviation is of this size in each of the eight directions, not only on average across them (Fig.~\ref{fig:fidelity}); averaged over the readings of each pair, it is 0.07 to 0.10~$\Delta E_{00}$ with a maximum of 0.45.
It is of the order of the minimal step defined in Section~\ref{subsec:ps_quantization}.
Color differences from threshold to a few units are therefore preserved through the drive, the rendering and several minutes of continuous operation.

\label{subsec:stability}%
The stability of the output over ten-minute intervals, with feedback enabled, was characterized by \citet{aksentsev2026colorimeter}.
The pair-separation measurements above were taken during continuous alternation, so they also show that the set differences are preserved while the stimuli alternate.
Over the 54-hour experiment the fourteen block means of the center error stay within 0.19 to 0.38~$\Delta E_{00}$ of each other at every center, with a slow linear drift of 0.06 to 0.22~$\Delta E_{00}$ over the experiment, and the block means of the pair-separation error within 0.01 to 0.07~$\Delta E_{00}$: the set differences are held more steadily than the centers themselves.
This covers the between-session stability that a multi-day study depends on.
\section{Configuring the colorimeter for the task}\label{sec:primary_selection}

The validation by \citet{aksentsev2026colorimeter} and in Section~\ref{sec:approbation} covered the instrument in its working three-primary configurations but did not address its main degree of freedom: the colorimeter can reproduce a given target with many different sets of calibrated wavelengths.
By definition, within the device model every feasible set yields the same tristimulus values for the standard observer and therefore matches the target equally well.
Reproduction accuracy therefore cannot distinguish between these alternatives, and no measurement of the kind reported in Section~\ref{sec:approbation} can favor one set over another.
The choice must instead be based on the psychophysical task for which the stimulus is intended.
This choice matters more than it might seem: depending on the color center, the feasible sets differ by a factor of three to eight in observer metamerism, measured as the mean distance of individual observers' percepts from the specified color.
Each task places its own demands on the stimulus, and each demand defines a criterion for selecting among the feasible sets.
In this section, the degrees of freedom described in Section~\ref{sec:freedoms} are fixed for each task in turn, and each criterion is developed together with the corresponding experimental setup.
All proposed configurations are implemented through the calibrated device model, whose accuracy at the working configurations was verified by \citet{aksentsev2026colorimeter} and in Section~\ref{sec:approbation}.
The section begins with the measurement of color-matching functions (Section~\ref{subsec:ps_stock}), where the criterion is the resolution of the quantized drive.
Color-difference measurement follows (Section~\ref{subsec:task_difference}), where the criterion is observer metamerism and a reference display serves as the baseline.
The section concludes with setups for flicker-induced colors and other tasks (Sections~\ref{subsec:task_fechner} and~\ref{subsec:impl_uses}).

\subsection{Setup for measuring color-matching functions}
\label{subsec:ps_stock}
\label{subsec:ps_quantization}

Color-matching functions are classically measured by having an observer match a test light with a mixture of adjustable narrowband primaries \citep{wright1928trichromatic,guild1931colorimetric,stiles1959npl}.
In such an experiment, how finely the observer can control the stimulus is limited by the adjustment step of the drive.
The selection criterion for this setup is therefore the color resolution of the quantized drive, that is, how finely the instrument can change the color it produces.
This is the natural criterion for a stimulus generator, whose purpose is to place colors precisely.

For each target, the selection routine searches all combinations of calibrated frequencies and discards those that cannot reproduce it.
A combination is feasible if its drive solution has non-negative intensities within each channel's calibrated range and uses primaries separated by at least 20~nm \citep{aksentsev2026colorimeter}, and if it also reproduces the eight colors at $\Delta E_{00}=3$ from the target, equally spaced in direction at the target's luminance, so that the neighborhood a color-difference experiment presents is reachable with the same primaries.
Among the feasible primary sets $\Lambda$, the routine selects the one whose output color is least sensitive to drive quantization, by minimizing
\begin{equation}\label{eq:ps_objective}
J(\Lambda)\;=\;\Delta E\bigl(\Vec{t},\,\Vec{C}(\hat{\Vec{P}}_{t})\bigr)\;+\;\max_{i,\;s=\pm1}\,\Delta E\bigl(\Vec{C}(\hat{\Vec{P}}_{t}),\,\Vec{C}(\hat{\Vec{P}}_{t}+s\,\delta P\,\Vec{e}_{i})\bigr),
\end{equation}
where $\Vec{t}$ is the target, $\hat{\Vec{P}}_{t}$ is the vector of channel drive amplitudes that reproduces $\Vec{t}$ with the set $\Lambda$, rounded to the drive quantum $\delta P$ (0.1~\% of full-scale amplitude, the least significant bit of the 10-bit command), $\Vec{e}_{i}$ is the unit vector along channel $i$, $s$ is the direction of the step, and $\Vec{C}(\cdot)$ is the color produced according to the calibrated device model.
With three primaries the drive solution is unique (Eq.~\eqref{eq:ps_coef3}).
With four, the solutions form the one-parameter family of Eq.~\eqref{eq:ps_coef4}, and the resolution solver picks its member by the second term of Eq.~\eqref{eq:ps_objective}, sweeping the family for the smallest single-step color change.
In Eq.~\eqref{eq:ps_objective}, $\Delta E$ denotes a color-difference formula.
The procedure does not depend on which formula is used, although the selected sets may.
This work uses CIEDE2000, $\Delta E_{00}$ \citep{luo2001development}, as in Section~\ref{sec:approbation}.

The first term is the placement error, the deviation of the produced color from the target caused by rounding the drive amplitudes to the quantum.
The second is the minimal step, the largest color change that a single drive quantum on one channel can produce.
Both terms are evaluated through the calibrated device model and enter with unit weight.
The minimal step is taken as the largest over the target and the eight neighborhood colors.
The working configurations of Section~\ref{sec:approbation} are three-primary sets selected on the calibration of the validation experiment by Eq.~\eqref{eq:ps_objective} with the minimal step evaluated at the target alone, among the sets that render the neighborhood ring of radius 5~$\Delta E_{00}$ (radius 3 at blue).
Before quantization, every feasible set reproduces the target exactly within the device model, so calibration accuracy acts only as a constraint on which sets are admissible.
What differs between sets is the error introduced by quantization, which both terms measure.

A matching observer cannot adjust the stimulus more finely than the minimal step, so it must lie well below the discrimination threshold.
Table~\ref{tab:criteria_accuracy} (Section~\ref{subsec:ps_choice}) lists the configurations this criterion selects.

\subsection{Setup for color-difference measurement}
\label{subsec:task_difference}

A color-difference experiment presents pairs of stimuli at threshold scale to real observers, none of whom is the standard observer, and its results must be comparable with existing color-difference data.
Stimuli that match for the standard observer do not match for individual observers, so each observer perceives a color displaced from the one the experiment specifies.
This observer metamerism is a property of the instrument and the observer population alone, but its magnitude cannot be interpreted in isolation, because a broadband display does not place its observers exactly on the specified color either.
The reference display of Section~\ref{sec:approbation} therefore serves as a baseline: an instrument intended to replace it should perform at least as well on the same measure, for the same observers.

\subsubsection{Evaluating observer metamerism}
\label{sec:metamerism}

Both devices reproduce the color centers in the same device-independent coordinates, so within their device models both stimuli match the specified colors for the CIE 1964 standard observer.
The display's measured residuals are given in Table~\ref{tab:accuracy}.
This match does not extend to individual observers.
By the mechanism of Section~\ref{sec:related}, a narrowband device is more exposed to observer differences than a broadband one, because its primaries sample each observer's color-matching functions at essentially one wavelength rather than averaging over a band.
How strongly this shows depends on which primaries are used.

Observer metamerism is evaluated for the 151 color-normal observers of the individual-observer database \citep{asano2016individual}, with the model evaluated for a $10^\circ$ field like the standard observer, each adapted to a field with the spectrum of CIE illuminant D50 at 100~cd/m$^2$.
For each observer, the departure is the $\Delta E_{00}$ \citep{luo2001development} between the color that observer perceives and the color the standard observer perceives, which is the specified color.
Each is expressed in CIELAB computed with its own observer's color-matching functions, against the D50 adapting white as that observer sees it.
The same measure is applied to the colorimeter and to the display reproducing the same color, so that both are read on one scale.

$\Delta E_{00}$ is calibrated for the CIE standard observer, and CIELAB built from an individual observer's color-matching functions is formally a different color space, in which that calibration is not guaranteed.
The difference is therefore taken under one fixed convention: the individual observer and the standard observer are each adapted to the same D50 field as they see it, so that a stimulus identical to the adapting field is neutral for every observer and scores zero.
The measure remains a surrogate, since no color-difference formula for individual observers exists, and it is used comparatively, to rank primary sets and devices under one fixed convention.

The resolution criterion of Eq.~\eqref{eq:ps_objective} does not take observers into account, so nothing ensures that a configuration selected by it places individual observers close to the specified color.
The next subsection introduces a criterion based on observers alone.
The departures produced by both criteria are compared with those of the display in Section~\ref{subsec:ps_choice}.

\subsubsection{The observer criterion}
\label{subsec:ps_spread}

Each color center was then reproduced with every set of primaries the calibration makes available: 19 wavelengths give 969 three-primary and 3876 four-primary sets, and, with the four-primary coefficient freedom sampled, discarding candidates that require a negative contribution leaves 20\,000--28\,000 feasible set--coefficient candidates per center.
With four primaries, the intensities can still move along the family of Eq.~\eqref{eq:ps_coef4} without changing the color for the standard observer. 
With three, they are fixed by Eq.~\eqref{eq:ps_coef3}.
The criterion of this setup scores each set $\Lambda$ by the lowest population-mean departure of the individual percepts from the specified color that any of these intensities achieves:
\begin{equation}\label{eq:ps_observer}
M(\Lambda)\;=\;\min_{\substack{\Vec{c}\,\ge\,0\\ A_{\Lambda}\Vec{c}\,=\,\Vec{t}}}\;\frac{1}{O}\,\sum_{o=1}^{O}\Delta E_{00}\bigl(L_{o}(T_{o}\,\Vec{s}_{c}),\;\bar{L}(\Vec{t})\bigr),\qquad \Vec{s}_{c}=\sum_{i\in\Lambda}c_{i}\,\Vec{S}_{i},
\end{equation}
where $\Vec{c}=(c_{i})_{i\in\Lambda}$ are the intensities of the primaries and $A_{\Lambda}$ the matrix of their tristimulus values, both as in Section~\ref{sec:freedoms}, $\Vec{S}_{i}$ is the calibrated spectrum of primary $i$, $\Vec{s}_{c}$ is the resulting colorimeter spectrum, $\Vec{t}$ is the tristimulus vector of the specified color (Section~\ref{sec:approbation}), and $T_{o}$ contains the color-matching functions of observer $o$ of the $O=75$ training observers.
$L_{o}(\cdot)$ and $\bar{L}(\cdot)$ convert to CIELAB against the D50 adapting white as observer $o$ and as the standard observer see it, the convention of the preceding subsection.
The constraint is the metameric match of Section~\ref{sec:freedoms}, so every admissible $\Vec{c}$ reproduces the target exactly for the standard observer, and the criterion ranks only what accuracy cannot.
The set is selected by the mean of Eq.~\eqref{eq:ps_observer}, among the sets that reproduce the target and its $\Delta E_{00}=3$ neighborhood, the same feasibility condition as for the resolution criterion.
Within the selected set's coefficient family of Eq.~\eqref{eq:ps_coef4}, the adopted member is then the one that minimizes the departure of the \emph{worst} observer rather than the mean, which protects the tail of the population.

This criterion selects a configuration by minimizing a score over the observer population, so scoring the selected configuration on the same observers would report a quantity to which the selection had already been fitted.
The population is therefore split once, before any selection: 75 observers are used for selection and 76 for reporting, stratified by age so that both halves have the same age distribution (mean 40 years).
Every configuration in this section is selected on the training half, and every observer-dependent number reported from here on comes from the held-out half.
The resolution criterion of Eq.~\eqref{eq:ps_objective} does not use observers, so its figures are out of sample in any case.
The split makes the two criteria comparable.
The split costs little, because the optimum is broad enough that half of the observers locate it nearly as well as all of them.

\subsubsection{Selection results}
\label{subsec:ps_choice}

Because the primaries are switched electronically, each color center can use its own primary set.
Every selected set was checked through the instrument's render path.
The choice of set has a large effect on observers.
Across the feasible four-primary sets, each driven with the solver's default member (the one that maximizes the smallest intensity), the mean departure of the population from the specified color differs by a factor of three to eight between the best and the worst set, depending on the center.
At yellow, for example, it ranges from 0.92 to 5.2~$\Delta E_{00}$.
The best sets are not rare outliers: the standard deviation across sets is 26 to 48~\% of the mean.

\begin{table}[tbp]
\caption{The configurations selected by the two criteria and the reference display, compared on reproduction accuracy and minimal step.
The CIE 1964 columns give the residual for the standard observer, and the minimal step is the second term of Eq.~\eqref{eq:ps_objective}.
Both are computed through each device's model and reflect quantization only, the display at its effective 12-bit depth.
Bold marks the best of the three in each row for each quantity.
All values in $\Delta E_{00}$.}\label{tab:criteria_accuracy}
\begin{tabular*}{\textwidth}{@{\extracolsep\fill}lcccccc}
\toprule%
& \multicolumn{2}{c}{Resolution-selected} & \multicolumn{2}{c}{Observer-selected} & \multicolumn{2}{c}{Display} \\
\cmidrule{2-3}\cmidrule{4-5}\cmidrule{6-7}
Center & CIE 1964 & Min.\ step & CIE 1964 & Min.\ step & CIE 1964 & Min.\ step \\
\midrule
Gray   & \textbf{0.02} & \textbf{0.14} & \textbf{0.02} & 0.40 & 0.04 & \textbf{0.14} \\
Yellow & \textbf{0.01} & 0.09 & 0.08 & 0.28 & 0.03 & \textbf{0.04} \\
Green  & 0.05 & \textbf{0.08} & \textbf{0.03} & 0.48 & \textbf{0.03} & 0.09 \\
Blue   & \textbf{0.02} & \textbf{0.09} & 0.05 & 0.45 & 0.09 & 0.16 \\
Red    & 0.02 & \textbf{0.13} & 0.02 & 0.18 & \textbf{0.01} & 0.14 \\
\botrule
\end{tabular*}
\end{table}

Table~\ref{tab:criteria_accuracy} compares the two selected configurations with the display on two quantities that do not require an observer model: the residual for the standard observer and the minimal step.
Residuals are 0.09~$\Delta E_{00}$ or less for all three, and cannot tell the generators apart.
Minimal steps range from 0.04 to 0.28~$\Delta E_{00}$, several times below typical discrimination thresholds, except for the observer-selected configurations at gray, green and blue (0.40 to 0.48~$\Delta E_{00}$): these drive a 510 or 543~nm primary at low amplitude, where the calibrated intensity curve is steepest, and the observer criterion does not see the step.
On minimal step the resolution criterion wins at every center, as it is built to.

\begin{table}[tbp]
\caption{The configurations selected by the two criteria and the reference display, compared on observer metamerism.
The mean and worst columns give the departure of the 76 held-out observers' percepts from the specified color, each in that observer's own CIELAB, averaged over the population and for the worst single observer.
Both are computed from each device's spectrum for the specified color, the observer-selected configurations at their adopted drive.
Bold marks the best of the three in each row for each quantity.
All values in $\Delta E_{00}$.}\label{tab:criteria_mismatch}
\begin{tabular*}{\textwidth}{@{\extracolsep\fill}lcccccc}
\toprule%
& \multicolumn{2}{c}{Resolution-selected} & \multicolumn{2}{c}{Observer-selected} & \multicolumn{2}{c}{Display} \\
\cmidrule{2-3}\cmidrule{4-5}\cmidrule{6-7}
Center & Mean & Worst & Mean & Worst & Mean & Worst \\
\midrule
Gray   & 3.06 & 9.58 & \textbf{1.11} & \textbf{2.90} & 2.38 & 6.93 \\
Yellow & 3.97 & 12.43 & \textbf{0.64} & \textbf{1.91} & 1.22 & 4.60 \\
Green  & 4.85 & 14.41 & \textbf{1.94} & \textbf{7.33} & 2.49 & 7.34 \\
Blue   & 3.65 & 9.58 & \textbf{1.34} & \textbf{4.60} & 2.52 & 7.80 \\
Red    & 1.88 & 6.66 & \textbf{1.24} & \textbf{4.50} & 1.37 & 4.56 \\
\botrule
\end{tabular*}
\end{table}

Table~\ref{tab:criteria_mismatch} shows how far the held-out observers' percepts fall from the specified color.
Here the generators differ.
With the resolution-selected configurations, the mean departure is 1.9 to 4.9~$\Delta E_{00}$, and individual observers depart by up to 14.4.
Since discrimination thresholds are of the order of 1~$\Delta E_{00}$, many observers see these stimuli several thresholds away from the intended color.
This is worse than the display (1.2 to 2.5) at every center.
With the observer-selected configurations, the mean departure falls to between 0.64 and 1.94~$\Delta E_{00}$.
This is 1.5 to 6.2 times less than with the resolution-selected configurations, and less than with the display at every center.
For the worst observer, they are also better than the display at every center, marginally so at green and red.

\begin{figure}[!htbp]
\centering
\includegraphics[width=\linewidth]{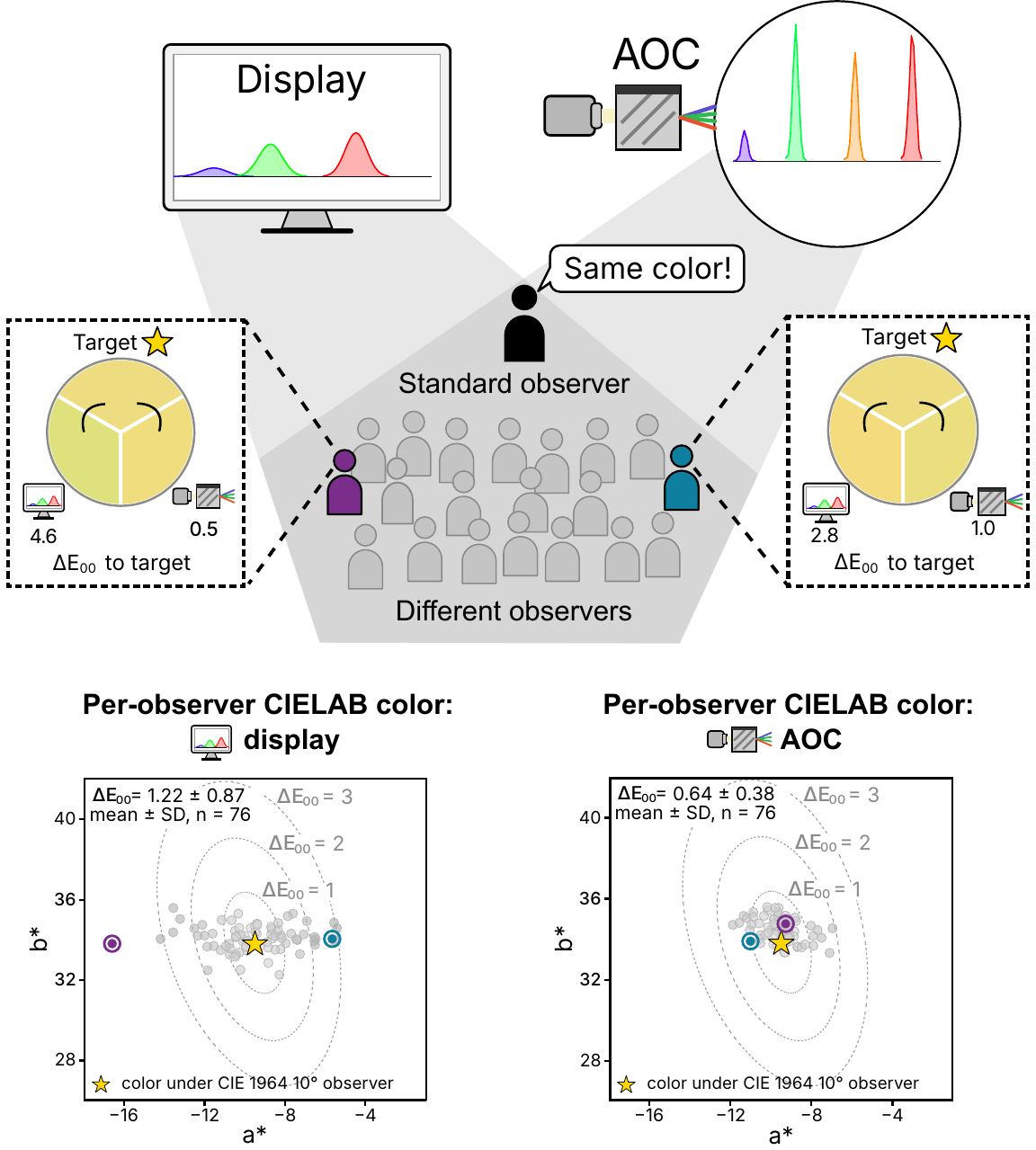}
\caption{The observer-metamerism mechanism at the yellow color center.
\textbf{Top}: two stimuli metameric for the CIE 1964 $10^\circ$ standard observer, the display rendering the center (schematic) and the observer-selected configuration at its adopted drive (its four bands, 443, 510, 587 and 660~nm, drawn to scale).
\textbf{Middle}: the standard observer sees the same color from both; individual observers do not. Two of them, marked in purple and teal, are traced to their percepts below, and the boxes give the departure of each from the specified color under the display and under the colorimeter.
\textbf{Bottom}: the color each of the 76 held-out observers perceives from each stimulus, as $a^{*}b^{*}$ in that observer's own CIELAB against the adapting white as that observer sees it. The dashed curves are the $\Delta E_{00}=1$, 2 and 3 loci about the specified color for the standard observer at equal lightness.}\label{fig:observer_spread}
\end{figure}

Figure~\ref{fig:observer_spread} illustrates the effect at yellow.
Both stimuli match the specified color for the standard observer, but individual observers see them differently.
With the observer-selected configuration the percepts cluster more tightly around the specified color, and the worst observer is 1.9~$\Delta E_{00}$ away, against 4.6 with the display.

Two factors explain the gain.
The first is the fourth primary.
At gray and yellow, the two brightest centers, the best three-primary sets give mean departures of 1.66 and 1.25~$\Delta E_{00}$, while the adopted four-primary sets give 1.11 and 0.64~$\Delta E_{00}$.
At green and blue the fourth primary improves the mean by 0.27 and 0.12~$\Delta E_{00}$.
At red the best three-primary set is level with the adopted four-primary one, 1.16 against 1.24~$\Delta E_{00}$.
Under the resolution criterion, by contrast, the fourth primary only reduces the minimal step and does not help observers.
The second factor is the drive amplitudes.
With four primaries, a one-parameter family of amplitude combinations matches the standard observer (Eq.~\eqref{eq:ps_coef4}).
At yellow, the adopted combination lowers the mean departure from 0.92 to 0.64~$\Delta E_{00}$ and the worst observer from 3.46 to 1.91~$\Delta E_{00}$, compared with the solver's default combination.
The stored amplitudes should therefore be used as they are, not recomputed by the solver.

\label{subsec:ps_weighing}%
Which criterion to use depends on the experiment.
For color-matching experiments, the resolution criterion is the right one: the minimal step limits how finely the observer can adjust the stimulus, and the observer's own departure from the standard observer is what the experiment measures.
For color-difference experiments, primaries should be selected by the observer criterion of Eq.~\eqref{eq:ps_observer}.
Where a finer minimal step is needed, for example by the stabilization loop, an upper bound on it can be added to the selection as a constraint.
Lower observer metamerism means that each participant sees a color closer to the specified one, so the measured thresholds refer to the intended color center.
This makes results obtained with the colorimeter directly comparable with color-difference datasets collected at the same centers on other devices.

A smaller minimal step remains an advantage in practice.
The online calibration (the stabilization loop of Section~\ref{sec:instrument}) corrects the output in increments of one drive quantum \citep{aksentsev2026colorimeter}, so a smaller minimal step lets it hold the color closer to its target during a session.
This gain is limited by the size of the step itself, which is already below threshold, whereas any departure from the specified color adds directly to the error of the experiment.
On this instrument the balance favors the observer criterion: the resolution criterion would reduce the minimal step by at most 0.40~$\Delta E_{00}$, but would increase the mean departure by up to 3.3~$\Delta E_{00}$.

These results have three limitations.
First, all configurations were evaluated with the calibrated device model.
The hardware measurements of \citet{aksentsev2026colorimeter} and of Section~\ref{sec:approbation} cover only the working configurations, which do not include 443~nm, a primary used by four of the five adopted sets.
Second, each center uses its own configuration, so results should be reported per center rather than pooled across centers.
Third, the five centers do not include highly saturated colors, where observer metamerism is expected to be largest (Section~\ref{subsec:lim_scope}).

\subsection{Setup for flicker-induced colors}
\label{subsec:task_fechner}

Achromatic stimulus that changes over time can produce colors that its spectrum alone does not predict.
Intermittent stimulation at a few hertz to a few tens of hertz produces the subjective colors described by \citet{fechner1838scheibe} and made widely known by Benham's top \citep{benham1894artificial}, which have since been studied as pattern-induced flicker colors \citep{campenhausen1995hundred}.
The classical stimulus is a spinning disc.
Its temporal pattern is fixed by the printed artwork and its spectrum by the illuminant, so neither can be varied independently.
The colorimeter removes both constraints.
Its acoustic drive switches the entire spectral composition electronically at rates up to a kilohertz, far above the frequencies at which the effect occurs, so each phase of the temporal pattern can be any calibrated spectrum.
This allows three configurations.
In the first, the phases are arbitrary chromaticities under full spectral control, which turns the classical demonstrations into parametric stimuli.
In the second, the phases are members of the metamer family of Eq.~\eqref{eq:ps_coef4}.
Such an alternation is invisible to the standard observer, so its visibility to an individual observer measures that observer's departure from the standard observer at the alternation rate.
In the third, the phases are different narrowband lines switched faster than the eye can follow, so that the observer integrates them into a single broadband stimulus.
This makes it possible to approximate broadband spectra, such as those of display primaries or daylight, on the same instrument, and so to compare narrowband and broadband stimuli without changing the device.
Section~\ref{sec:approbation} verified the stimuli only under slow alternation, with a cycle of 0.2~s that includes the spectroradiometer capture.
The fast switching used here, up to 1~kHz, has not been measured, so this setup is a configuration proposal rather than a validated protocol.

\subsection{Further setups}
\label{subsec:impl_uses}

Four more configurations rest on the same machinery and are likewise stated as proposals.

\begin{itemize}
\item \textbf{Photoreceptor-directed stimulation by silent substitution.}
Oscillating along the $\tau$ of Eq.~\eqref{eq:ps_coef4} rather than fixing it yields stimuli identical for three receptor classes and distinct for a fourth, such as melanopsin \citep{spitschan2018silent}.
Three provisos apply.
The silenced quantities must be cone fundamentals rather than CIE 1964 tristimulus values, and individual departures from the standard observer must be corrected for, because otherwise more sensitive cone processes intrude on a nominally melanopsin-directed stimulus \citep{uprety2021optimizing}.
Rod intrusion must be suppressed by working photopically.
Finally, the $2^\circ$ field is small for melanopsin protocols.
The population evaluation of Section~\ref{subsec:task_difference} then gives what this field most needs to bound, the distribution of silencing failure across real observers, and the selection that minimizes observer metamerism also minimizes this leak.
\item \textbf{Observer classification.}
The search that minimizes Eq.~\eqref{eq:ps_observer} can as cheaply maximize it, producing pairs that match for the standard observer while separating real observers as widely as the gamut allows: an efficient probe for classifying participants by their departure from the mean observer, and a direct empirical test of the observer model itself.
\item \textbf{Objective colorimetry on animals.}
The substitution principle this instrument implements electronically is the one Bongard used to perform colorimetry without a verbal report: alternating two lights and finding the pair for which a recorded response vanishes, on the frog's optic nerve \citep{bongard1955animals} before the psychophysical form of the method was published \citep{bongard1957substitution}.
What the alternation asks is the same question Section~\ref{subsec:task_difference} asks of the observer population -- whether two spectra are equivalent for a given set of receptors -- with a neural or behavioral response in place of a match.
Four primaries silence three receptor classes while modulating a fourth, so for a tetrachromatic animal, as birds and many fish are, a single cone class can be driven alone.
The isolation is then exactly determined, with no coefficient freedom left to trade against other constraints.
Species whose short-wave cone lies in the ultraviolet are out of reach of the present calibration, which spans 443 to 660~nm, though not of the source, which reaches 380~nm.

\item \textbf{Characterization of devices that stand in for observers.}
A tunable narrowband source measures a camera's spectral sensitivities directly, the monochromator-based measurement against which indirect estimates are judged \citep{tominaga2021measurement}, and observer-metameric pairs maximally distinct for the sensor measure the camera--observer disagreement that a calibration for human-facing imagery has to bound.
\end{itemize}
\section{General discussion}\label{sec:discussion}

Nothing in the preceding argument depends on acousto-optics.
It requires only that the instrument build its stimulus from a small number of narrow spectral bands, and that more than one combination of those bands can reach a given target.
Multi-primary LED systems, laser projectors and monochromator-based generators satisfy both conditions.
The recommendations below therefore apply to this class of instruments, and so do the limitations that follow them.

\subsection{Recommendations for users of narrowband generators}\label{subsec:impl_checklist}

Together with \citet{aksentsev2026colorimeter}, Sections~\ref{sec:approbation} and~\ref{sec:primary_selection} amount to five properties that should be established before observer studies with a narrowband generator.

\begin{enumerate}
\item Accuracy at the operating points actually used, measured under the conditions of the experiment \citep{aksentsev2026colorimeter}.
\item Accuracy of the color differences the experiment presents, measured with the stimuli alternating as in a trial (Section~\ref{sec:approbation}).
\item Stability at the timescale of a trial and of a session (\citealp{aksentsev2026colorimeter}; Section~\ref{sec:approbation}).
\item Uniformity over the field an observer integrates \citep{aksentsev2026colorimeter,aksentsev2026homogenization}.
\item Observer metamerism, the departure of individual observers' percepts from the specified color, computed for a published observer population and compared with a baseline device (Section~\ref{sec:primary_selection}).
\end{enumerate}

The last property is routinely omitted, because an instrument that reproduces its targets for the standard observer has done what colorimetry asks of it.
It also differs from the other four in kind.
It is not a check applied to a configured instrument but a criterion for configuring it, and reproduction accuracy cannot detect it, since every feasible primary set is equally accurate.
The remedy costs no light, hardware or accuracy.
The equivalent sets are ranked on the observer criterion, a computation over calibration data that needs no additional measurement.
On the instrument studied here, this reduces the mean departure by a factor of 1.5 to 6.2, to below that of a calibrated display at every color center (Section~\ref{subsec:ps_choice}).

\subsection{Limitations}\label{subsec:lim_observers}
\label{subsec:lim_prediction}%
\label{subsec:lim_scope}%
\label{subsec:lim_conditioning}%
The observer results of Section~\ref{sec:primary_selection} are computed rather than measured.
They rest on measured spectra, a published observer model \citep{asano2016individual} and standard colorimetry, and the primaries were selected on 75 observers and evaluated on the other 76.
Four limitations follow.
First, the model population contains only the variability its parameters can express, and the primary spectra are recorded broader than they are (Section~\ref{sec:instrument}), so the departures reported here are lower estimates, and the results are comparative rather than predictions for individual observers.
Second, the selection depends on the observer data it is based on: configurations selected for the CIE 2015 physiological observer, the two Stiles--Burch observers or the model evaluated over a range of ages transfer to the observers reported here at a cost of 0.6 to 1.6~$\Delta E_{00}$, so a laboratory serving a different population should repeat the selection rather than reuse these wavelengths.
Third, no experiment with observers has been carried out yet, and of the configurations proposed here only the three-primary resolution-selected sets of Section~\ref{sec:approbation} have been measured on hardware; the four-primary configurations of Tables~\ref{tab:criteria_accuracy} and~\ref{tab:criteria_mismatch}, including all observer-selected ones, have not.
Fourth, the adapting field is modeled by the tabulated spectrum of CIE illuminant D50.
A real adapting field with a different spectrum gives each observer a different white, so the measured spectrum of the field should replace the table before the selection is used with real observers.

The analysis also covers only the five CIE color centers, with a single display as the baseline.
The saturated region, where the instrument is most useful, remains to be examined.
\section{Conclusion}\label{sec:conclusion}

This paper took the acousto-optic colorimeter described by \citet{aksentsev2026colorimeter} from characterization to use in color psychophysics.
Building on the accuracy, stability and uniformity reported there, it showed that the instrument reproduces what a color-difference experiment depends on: set color differences of 0.4 to 3.7~$\Delta E_{00}$, presented in alternation, were realized with a mean absolute error of 0.08 to 0.12~$\Delta E_{00}$ at all five CIE color centers.

Because the same color can be produced by many sets of primaries, all equally accurate for the standard observer, the instrument must also be configured for the task.
For color-difference measurement, the choice matters for individual observers.
With primaries selected only for drive resolution, the percepts of 151 model observers depart from the specified color by more than with a calibrated display.
With primaries selected to minimize this departure at each color center, the mean departure falls by a factor of 1.5 to 6.2, below the display at every center, at no cost in light, hardware or reproduction accuracy.
For the measurement of color-matching functions, the smallest color step of the drive is the better criterion.
Further applications, from flicker-induced colors to silent substitution and observer classification, are stated as proposals.

Two conclusions apply beyond this device to any generator built from a small number of narrow primaries.
Reproduction accuracy for the standard observer is not a sufficient criterion for configuring such an instrument, because the standard observer is an average and experiments are run on individuals.
The resulting observer metamerism can, however, be reduced by selecting the primaries for the observer population, a computation from calibration data made before the first observer is seated.
On the instrument studied here, this reduces the mean departure from 1.9--4.9 to 0.64--1.94~$\Delta E_{00}$, against 1.2--2.5~$\Delta E_{00}$ for a calibrated display.
Because the selection is fitted to an observer population, a laboratory serving a different population should repeat it rather than reuse the wavelengths reported here.
The selected configurations were evaluated through the calibrated device model; on hardware, the three-primary resolution-selected sets have been verified (Section~\ref{sec:approbation}), and the four-primary configurations remain to be verified.
\section*{Declarations}

\begin{itemize}
\item \textbf{Funding.} The research was carried out within the state assignment, approved by the Ministry of Science and Higher Education of the Russian Federation (theme No. FFNU-2025-0045).
\item \textbf{Conflict of interest/Competing interests.} The authors declare no competing interests. 
\item \textbf{Ethics approval and consent to participate.} Not applicable. 
\item \textbf{Consent for publication.} Not applicable.
\item \textbf{Availability of data and materials.} The calibration of the colorimeter, the spectra of the reference display and its primaries, and the validation measurements of both devices analyzed in this study are publicly available at \url{https://github.com/createcolor/aoc-color-psychophysics}. The individual observer data used in Section~\ref{sec:primary_selection} are publicly available as part of the dataset of \citet{asano2016individual}.
\item \textbf{Code availability.} The code that reproduces the validation results and the primary selection of this study is publicly available at \url{https://github.com/createcolor/aoc-color-psychophysics}.
\item \textbf{Open practices statement.} The data and code for all analyses are available at \url{https://github.com/createcolor/aoc-color-psychophysics}.
\end{itemize}

\bibliography{sn-bibliography}

\end{document}